\documentclass[Universe,article,accept,moreauthors,pdftex,10pt,a4paper]{Definitions/mdpi} 

\firstpage{1} 
\pubvolume{xx}
\issuenum{1}
\articlenumber{5}
\pubyear{2018}
\copyrightyear{2018}
\Title{The Pierre Auger Observatory: review of latest results and perspectives}

\Author{Dariusz Gora $^{1}$*\orcidA{} for the Pierre Auger Collaboration  $^{2,\dagger,\ddagger}$}

\AuthorNames{Firstname Lastname, Firstname Lastname and Firstname Lastname}

\address{%
$^{1}$ \quad Institute of Nuclear Physics Polish Academy of Sciences, Radzikowskiego 152,
Cracow, Poland; Dariusz.Gora@ifj.edu.pl \\
$^{2}$ \quad Observatorio Pierre Auger, Av. San Martín Norte 304, 5613 Malargue }

\corres{Correspondence: Dariusz.Gora@ifj.edu.pl}

\firstnote{auger\_spokespersons@fnal.gov} 
\secondnote{Full author list: http://www.auger.org/archive/authors\_2018\_07.html}
\abstract{The Pierre Auger Observatory is the world's largest operating  detection system for the observation of ultra
high energy cosmic rays (UHECRs). The detector allows
 detailed measurements of their  energy spectrum, mass composition and
arrival directions of primary cosmic rays in the energy range above
$10^{17}$ eV. The data collected at the Observatory over the last decade
show the suppression of the cosmic ray flux at energies above $4\times10^{19}$
eV. However, it is still unclear if this suppression is caused by
the propagation of cosmic rays or rather by energy limitation of their
sources. The other puzzle is the origin of UHECRs. Some clues can be
drawn from studying the distribution of their arrival directions. The
recently observed dipole anisotropy has an orientation which indicates
an extragalactic origin of UHECRs. The Auger surface detector array is also
sensitive to showers due to ultra high energy neutrinos of all flavours and photons, and 
recent neutrino and photon limits provided by the Observatory
can constrain models of the cosmogenic neutrino production and 
exotic scenarios of the UHECRs origin, such as the decays of super
heavy particles. In this paper the recent results on measurements of the energy
spectrum, mass composition and arrival directions of cosmic rays, and future prospects are presented.}


\begin{document}
%%%%%%%%%%%%%%%%%%%%%%%%%%%%%%%%%%%%%%%%%%
%% Only for the journal Gels: Please place the Experimental Section after the Conclusions

%%%%%%%%%%%%%%%%%%%%%%%%%%%%%%%%%%%%%%%%%%
%\setcounter{section}{-1} %% Remove this when starting to work on the template.
%\armaceutics,polymers,processes,technologies,viruses,vision

\section{Introduction}
\label{intro}
The detection of ultra high energy cosmic rays (UHECRs), above $10^{17}$ eV, is important as it may
allow us to answer some of the most important questions in astrophysics: what are the sources of
high energy cosmic rays (CRs), where are they produced, what are the corresponding acceleration mechanisms, and what is their elemental composition. In fact, although the existence of UHECRs has been experimentally proven for at least 50 years, these questions are still open, mainly due to the small flux  of UHECRs which is one of the reasons for the slow progress in their understanding. The detection of CRs is also important because UHECRs 
may allow us to probe particle physics at an  energy scale beyond TeV energies. When CRs arrive at Earth, they collide with nuclei in the atmosphere at center-of-mass energies which are orders of magnitude above the ones available in man-made particle accelerators. CRs also propagate through the interstellar and intergalactic media and are thus subject to magnetic fields and to interactions with cosmic matter and radiation, therefore having the possibility to provide indirect indications about phenomena which arise only at large distances and highest energies.

Cosmic rays reaching the Earth's atmosphere interact with nuclei, creating a large number of lower energy particles in a showering
 process. These interactions  produce charged pions which can interact or decay, producing both atmospheric neutrinos and high energy muons. Neutral pions  decay  into gammas, feeding the so-called electromagnetic component of
the shower which  carries most of the initial energy. Only at the highest energies can showers induced by CRs reach the ground level and  be efficiently sampled by an array of surface detectors. Such  showers are commonly called Extensive Air Showers (EAS). The electromagnetic component of an EAS produces isotropic fluorescence light by exciting nitrogen molecules in the air. For very high
energies of the primary particle, enough fluorescence light is produced by the large number of secondaries in the cascading process so that the shower can be recorded from a distance of many kilometers by an appropriate optical detector system (fluorescence telescopes). As the amount of fluorescence light is correlated with energy dissipated by shower particles, it provides a calorimetric measure of the primary energy.

%The Pierre Auger Observatory~\cite{auger1, auger2}  combines a surface array (SD) to measure secondary particles at ground %level together with a fluorescence detector (FD) to measure the development of air showers in the atmosphere above the array.  %This hybrid detection technique enables  combining the calorimetric measurement of the
%shower energy through fluorescence light with them  high-statistics data of the surface array. 
In this paper after a brief description of the  Auger Observatory we report recent results from  the  Observatory i.e. the measurement of UHECR spectrum, the anisotropy  of arrival directions, and on the composition. Finally,  also limits on the UHE neutrino flux and limits on the flux of UHE photons are presented.
\begin{figure}[t]
\centering
\includegraphics[width=6.0 cm]{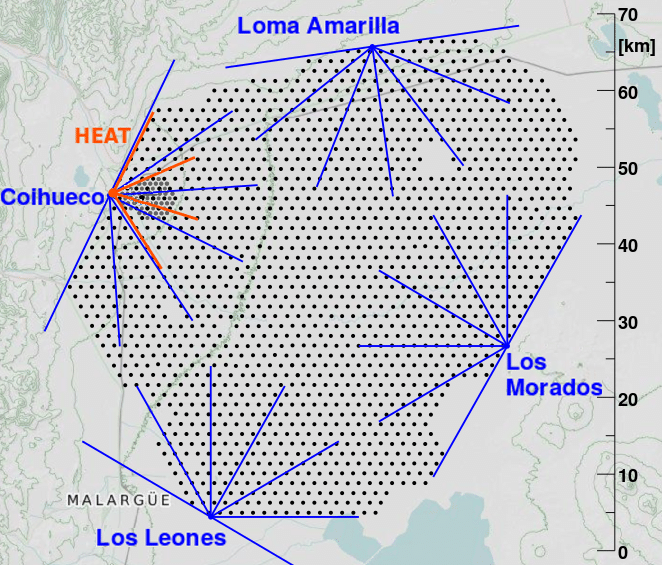}
\includegraphics[width=9.5 cm,height=6.5cm]{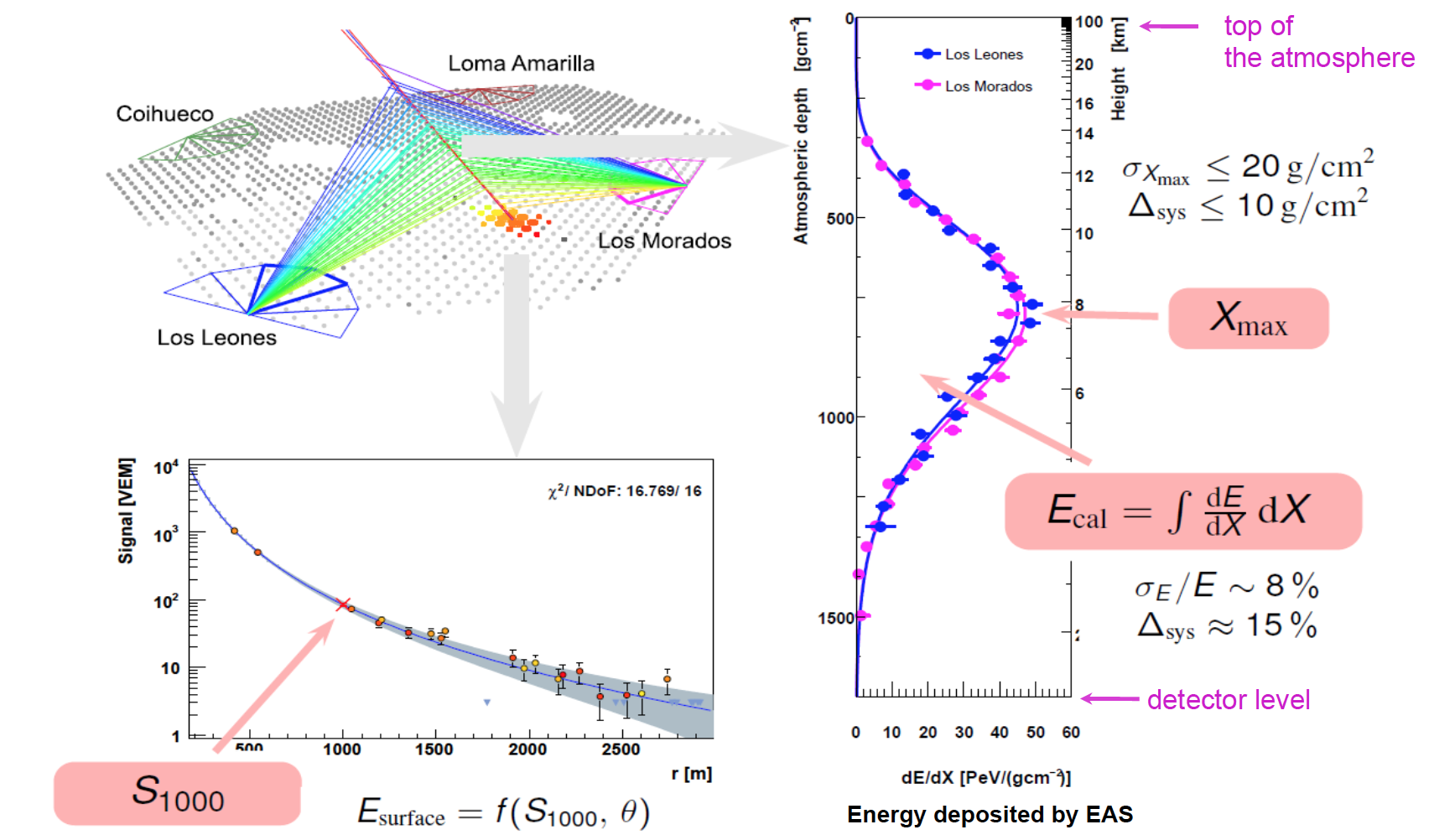}
\caption{  Left: Map of the Pierre Auger Observatory. The four telescope sites are marked in blue with lines indicating the field of view of the telescopes. The surface detector stations are represented by black dots. Right: Example of an event seen simultaneously by the SDs and FDs (hybrid event). The right plot inset shows the measured energy deposit profile of an EAS from the FD, the left plot shows the measured signal from stations of the SD with a fitted lateral distribution function (LDF). The LDF fit allows determining the particle density at a distance 1000 m from the shower axis.}
\label{fig-auger1}
\end{figure} 
\section{The Pierre Auger Observatory}
The Pierre Auger Observatory~\cite{auger1, auger2}   operating since 2004 is the largest project to measure UHECRs.
 It is located near Malargue in the province of Mendoza, Argentina. It combines a surface array (SD) to measure secondary particles at ground level together with a fluorescence detector (FD) to measure the development of air showers in the atmosphere above the array. Therefore this hybrid detection technique combines the calorimetric measurement of the shower energy through fluorescence light with the  high-statistics data of the surface array. 

The SD array consists of about 1600 water-Cherenkov detectors  arranged in a triangular grid with 1.5 km spacing (SD-1500 array)  covering an area of about 3000 km$^{2}$. In addition  61 detectors are distributed over  23.5 km$^{2}$ on a 750 m grid  (SD-750 or ‘infill’ array), see Figure ~\ref{fig-auger1} (left). Each water-Cherenkov detector has three photomultipliers (PMTs) on the top, which sample the shower signal. The signal detected at each station is expressed in  a common calibration unit, the so-called vertical equivalent muon or VEM~\cite{augerSD}. The SD  array  is able to collect EAS's at any time  with almost 
100 \% duty cycle.

The fluorescence detector (FD) consists of four telescope buildings (eyes) overlooking the detector array as it is shown in Figure~\ref{fig-auger1} (left). Each building houses 6 telescopes with a $30^{\circ} \times 28.6^{\circ}$ field of view.
The fluorescence light is focused in each telescope onto a camera consisting of 440 PMTs through its
 Schmidt-optics with a spherical mirror of  $\sim13$ m$^{2}$. The fluorescence detector probes the longitudinal  development of EAS (Figure~\ref{fig-auger1} (right))  by measuring the photon emission from atmospheric nitrogen, which is excited by shower charged particles. However, it can operate only during clear moonless  nights ($\simeq 15$  \% duty cycle). Three additional telescopes pointing at higher elevations (HEAT) are located near one of the FD sites (Coihueco) to detect lower energy showers. An  array of radio antennas, the Auger Engineering Radio Array (AERA)~\cite{area,auger2}, complements the data with the detection  of the shower radiation in the 30-80 MHz region. Details of the design and  status of the Observatory can be found in~\cite{auger1, auger2}.
 \section{Results}
 \subsection{High Energy Cosmic Ray Spectra}
\begin{figure}[t]
\centering
\includegraphics[width=6.0 cm]{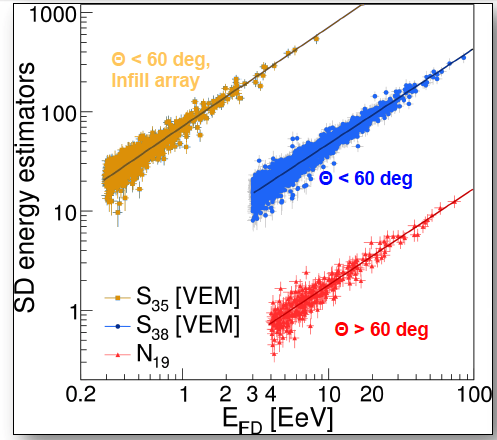}
\includegraphics[width=8 cm]{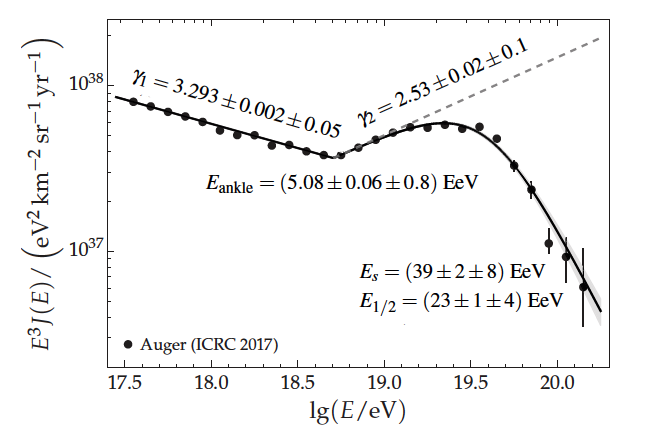}
\caption{ Left: Correlation between the SD energy estimators and the energy measured with the fluorescence telescopes ($E_{FD}$). The full line is the best
fit to data. Right: The Auger energy spectrum with two empirical fit functions (see Ref.~\cite{augerErrors}). The fitted spectral indices and energies of the break and suppression are superimposed together with their statistical and systematic uncertainties. }
\label{fig-cal}
\end{figure}

  The hybrid nature of the Auger Observatory enables us to determine the energy spectrum of primary CRs without strong dependence on our limited knowledge of the mass composition and hadronic interaction models. The Auger approach is to use a selected sample of hybrid events (see Figure~\ref{fig-auger1} (right)) in which the SD energy, $E_{SD}$ can be estimated  using the FD. The calibration curves, which are used to find energies of SD events are shown in Figure~\ref{fig-cal} (left). The parameter chosen to characterize the size of an SD event for the SD-1500 array is the signal at 1000 m from the shower axis, normalized to mean zenith angle of the events of $ 38^{\circ}$ ($S_{38}$) according to the method described in~\cite{augerCAL}. For  the SD-750 array the mean zenith angle of the events of $35^{\circ}$ ($S_{35}$) is used, while for inclined showers (with the zenith angle larger than $60^{\circ}$  seen by  SD-1500 array), the ($N_{19}$) energy estimator  is used~\cite{augerInclined}.  Following this method the overall systematic uncertainty of the energy scale remains at 14\%~\cite{augerErrors}.

In Figure~\ref{fig-cal} (right) the Auger combined energy spectrum as presented at ICRC 2017 is shown. The energy spectrum was 
obtained from  the four spectrum components  calculated for the hybrid events, events with the SD-1500/SD-750 array and  inclined events. The energy spectrum derived from hybrid data is combined with the one obtained from  SD  data using a maximum likelihood  method~\cite{augerCombained}. From the plot we can see some characteristic features of this  combined spectrum: the so-called “ankle” and the flux suppression at highest energies.
The position of the ankle at $ E_{ankle} = 5.08 \pm0.06(\mbox{stat.}) \pm0.8 (\mbox{syst.})$ EeV has been determined by fitting the flux with a broken power law $E^{-\gamma}$~\cite{augerErrors}. An index of  $\gamma= 3.293 \pm0.002(\mbox{stat.}) \pm0.05 (\mbox{syst.}$) is found below the ankle.  Above the ankle the spectrum follows a
power law  with index  $\gamma= 2.53 \pm0.02(\mbox{stat.}) \pm0.1 (\mbox{syst.)}$ with the suppression of the spectrum at $ E_{s} = 39 \pm 2(\mbox{stat.}) \pm 8 (\mbox{syst.})$ eV.   The energy at which the integral flux drops by a factor two below what would
 be expected without suppression is found to be $E_{1/2} = 23 \pm1(\mbox{stat.}) \pm 4 (\mbox{syst.}) $ EeV. This value is considerably lower than $E_{1/2} = 53 $ EeV as predicted for the classical GZK  scenario~\cite{GZK} in which the suppression at ultra high energies is caused by depletion of extra-galactic protons
 during propagation. The suppression of the spectrum can also be described by assuming a mixed composition at the sources or by the limiting acceleration energy at the sources rather than by the GZK-effect~\cite{GZKa,GZKb}.
\subsection{Composition}

The atmospheric depth $X_{\textnormal{max}}$ where an air shower reaches its maximum size (see as an example Figure~\ref{fig-auger1} (right))
depends on the primary nuclear mass. For a fixed energy, showers  initiated by 
heavy nuclei reach their maximum size on average at smaller $X_{\textnormal{max}}$ than those initiated by protons.  Shower-to-shower fluctuations in  $X_{\textnormal{ max}}$ are much less for heavy nuclei than for protons. Both of these properties are robust expectations that follow from a heavy
nucleus being composed of many nucleons. 

\begin{figure}[t]
\centering
\includegraphics[width=12.0 cm]{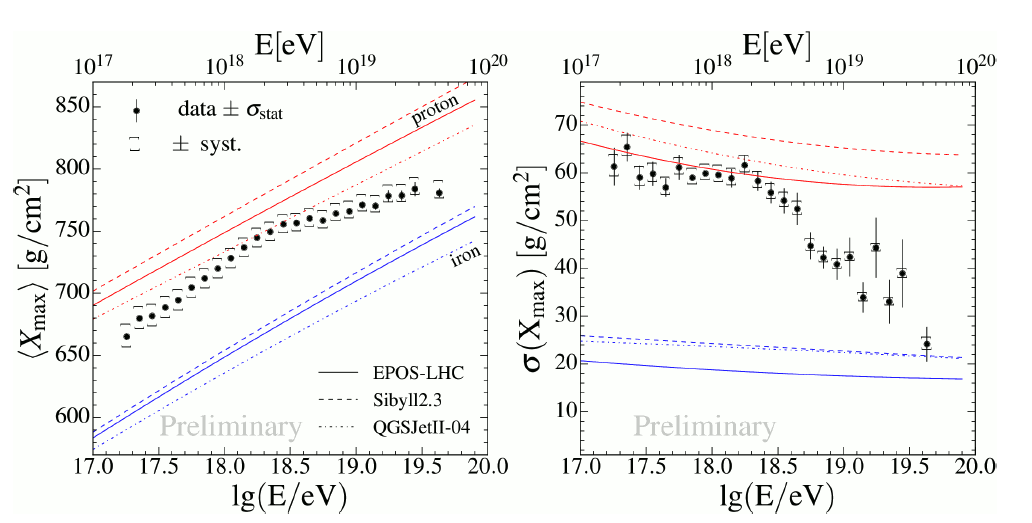}
\caption{ The mean $X_{\textnormal{max}}$ (left) and the standard deviation of the $X_{\textnormal{max}}$  (right) obtained by the FD as a function of the energy. The predictions of different
hadronic interactions models, for pure proton and iron composition, are also shown (see Ref.~\cite{augerXmax}) }
\label{fig-xmax}
\end{figure}

In Figure~\ref{fig-xmax} we show the measured  $ \langle X_{ \textnormal{max}}  \rangle$ and the shower-to-shower fluctuations, $\sigma(X_{\textnormal{ max}})$ from the Auger Observatory as also presented at ICRC 2017~\cite{augerXmax}. The linear fit, $\langle X_{\textnormal{ max}} \rangle= D_{10}  \cdot \lg(E/eV ) + c$, yields an elongation rate (variation of $ \langle X_{\textnormal{max}}  \rangle$ per decade of energy) of $D_{10}=79 \pm 1$ g/cm$^{2}$/decade
between $10^{17.2}$ and $10^{18.33}$ eV, see Figure~\ref{fig-xmax} (left).  This value, being larger than that expected for a constant mass composition ($\sim 60 $ g/cm$^{2}$/decade), indicates that the mean primary mass  becomes lower with increasing energy.  At $10^{18.33 \pm 0.02}$ eV the elongation rate becomes significantly smaller ($D_{10}=26 \pm 2$  g/cm$^{2}$/decade) indicating that the composition becomes heavier with increasing energy. Also the  $\sigma(X_{\textnormal{max}})$ shows a similar  behaviour, see Figure~\ref{fig-xmax}  (right). As it can be seen from the plot, the shower-to-shower fluctuations decrease  from 60 g/cm$^{2}$ to about 30 g/cm$^{2}$ as the energy increases. This decreasing fluctuations are an independent signature of  increasing average mass of the primary particles.

\begin{figure}[t]
\centering
\includegraphics[width=12.0 cm]{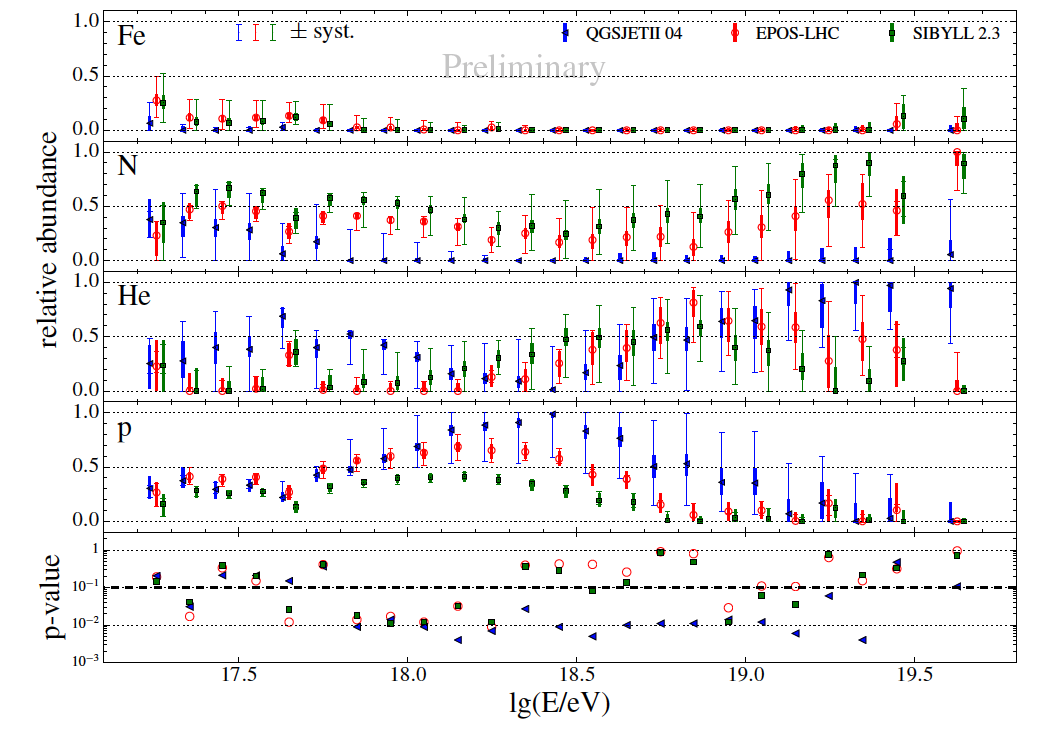}
\caption{ Mass fraction fits obtained using parameterizations of the $X_{\textnormal{max}}$ distributions from the
fluorescence  $X_{\textnormal{max}}$ data. The error bars indicate the statistics (smaller cap) and the systematic
uncertainties (larger cap). The bottom panel indicates the goodness of the fits (p-values), see ~\cite{augerXmax} for more details. }
\label{fig-xmax2}
\end{figure}

%%%%%%%%%%%%%%%%%%%%%%%%%%%%%%%%%%%%%%%%%%
The Auger $X_{\textnormal{max}}$ data (distributions  and their moments) enable a step further in the interpretation of mass composition by studying the evolution  of the first two
moments of $\ln A$ with energy~\cite{augerXmax,augerXmax2,augerXmax3} as well as the evolution of the fractions of four mass groups (H, He, N and Fe) obtained from the fit of the $X_{\textnormal{max}}$ distributions ~\cite{augerXmax,augerXmax2,augerXmax2a,augerXmax3}.
The latest results are shown in  Figure~\ref{fig-xmax2}. At the lowest energies, we find hints for a contribution 
(25-38 \% depending on  hadronic interaction model used) from iron primaries that disappears rapidly with increasing energy. At
 higher energies the composition is dominated by different elemental groups, starting from protons below the ankle and going through helium to nitrogen as the energy increases. This evolution occurs with apparently limited mass mixing as a consequence of the small values of the $X_{\textnormal{ max}}$ dispersion. The interpretation of $X_{\textnormal{max}}$ data depends on the hadronic interaction model. In particular QGSJETII-04 appears to be less consistent with data as can be seen in the lower panel of Figure~\ref{fig-xmax2}, where the p-value of the fit is shown.
This is because the present interpretation relies on air  shower simulations that use hadronic interaction
models to extrapolate particle interaction properties over two orders of magnitude in center-of-mass energy beyond the regime where they have been tested experimentally. The possible different interpretation can be that the proton-air interaction cross section or multiplicity, or both, are increased leading to a faster shower development, as  would happen in heavy nuclei collisions~\cite{augerXmax} . In other
words, the precise interpretation of the Auger $X_{max}$ and  $\sigma (X_{max})$ result is still an open question.

\subsection{Anisotropy }
An anisotropy in the arrival direction distribution of CRs in the energy range of the
GZK suppression is expected because of the highly anisotropic matter distribution on distance scales of
 100 Mpc. The  extragalactic  magnetic fields are not strong enough to deflect significantly high energy cosmic
rays during their propagation to the Earth, so the observed source distribution in principle should 
reflect the matter distribution in the nearby universe. The Auger Collaboration has undertaken several  anisotropy searches at different energy ranges and
angular scales. These use several tools like harmonic analysis, auto-correlation, correlation with 
source catalogs, and search for flux excesses in the visible sky and correlations with other experiments.

\begin{figure}[t]
\centering
\includegraphics[width=7.0 cm]{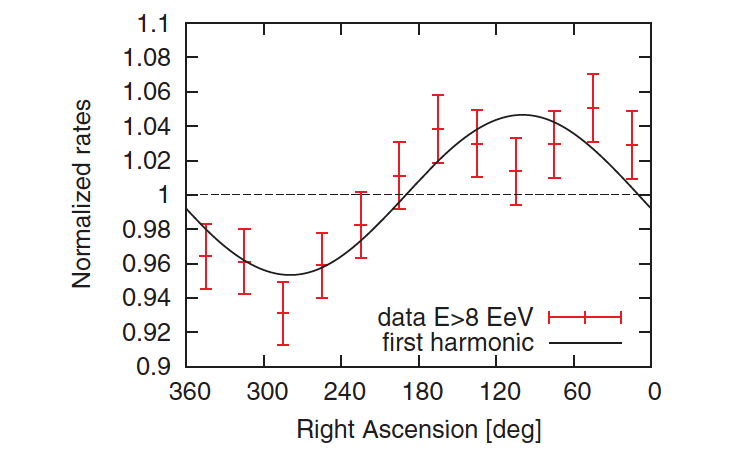}
\includegraphics[width=7.0 cm,height=5cm]{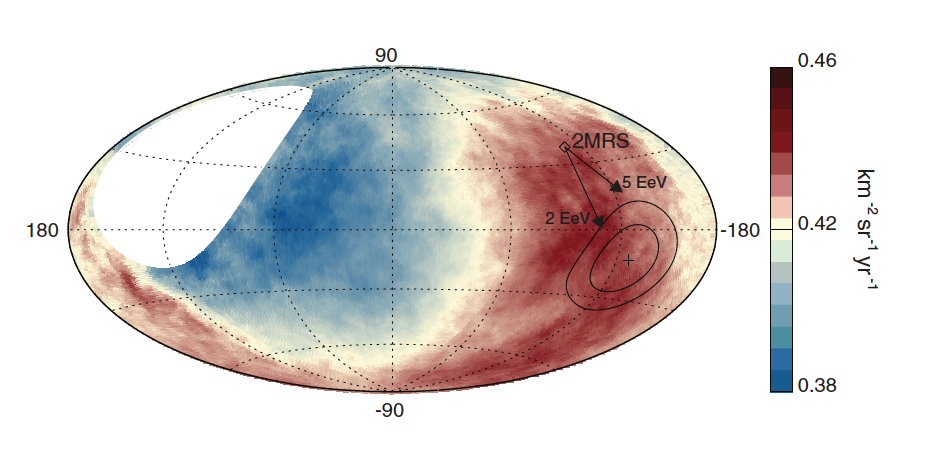}
\caption{ Left: Normalized rate of events as a function of right ascension for $E \geq 8$ EeV. The error bars represent $ 1 \sigma$ uncertainties~\cite{science}. Right: Sky map in galactic coordinates of the cosmic ray flux for $E \geq 8$ EeV. The cross indicates the dipole direction and the contours mark the 68 \% and 95 \% confidence level regions~\cite{science}. The diamond marks the dipole direction of the 2MRS galaxy distribution~\cite{2mrs}. The arrows show the deflections expected from the  model described in~\cite{science2} for particles with $E/Z= 5$ or 2 EeV. }
\label{fig-ani}
\end{figure}

Among the most recent studies the most exciting result is  the observation of a large-scale anisotropy in the arrival directions of CRs above $8 \times 10^{18}$ eV~\cite{science}. Two energy bins, 4 EeV$ <E \leq 8$ EeV  and  $E \geq 8$ EeV, were analyzed in data obtained since the start of data taking  with the Observatory  (total exposure of 76,800 km$^{2}$ sr yr) to  determine  the amplitude of the first harmonic in right ascension, Figure~\ref{fig-ani} (left). The right ascension anisotropy found in the two energy bins has amplitudes $0.5^{+0.6}_{-0.2}$ \% and $4.7 ^{+0.8}_{+0.7}$ \%, respectively. The  events in the lower energy bin follow an arrival direction distribution
consistent with isotropy, but in the higher energy bin a significant anisotropy was found, with a p-value of $2.6 \times 10^{-8}$ under the isotropic null hypothesis. The three-dimensional dipole, obtained by combining the first-harmonic analysis in right ascension with a similar one in the azimuthal angle, has a direction in galactic coordinates (l, b) = ($233^{\circ} $, $-13^{ \circ} $) which is  about 125$^{ \circ }$ away from the Galactic Centre.  Hence this anisotropy  indicates an extragalactic origin for these UHECR particles. The dipole anisotropy has an amplitude   of $6.5^{+1.3}_{-0.9}$\% and level of significance of 5.2 $\sigma$. A  skymap of  the intensity of CRs arriving above 8 EeV is shown in Figure~\ref{fig-ani} (right).

\subsection{Photons and neutrinos }
Another important issue concerning composition studies is the search for photons and neutrinos in
primary CRs. Photons and neutrinos may directly trace back the origin of the UHECR. Essentially all
models of UHECRs production predict neutrinos/photons as  a result of the decay of charged/neutral pions produced in
interactions of CRs within the sources themselves or while propagating through background 
radiation fields. For example, UHECR protons interacting with the cosmic microwave background (CMB) give rise to the so-called “cosmogenic” or GZK neutrinos~\cite{gzkneutrino}. The cosmogenic neutrino flux is somewhat uncertain since it depends on the primary UHECR composition and on the nature and cosmological evolution of the sources as well as on their spatial distribution. In general, about 1 \%  of cosmogenic neutrinos is expected in the ultra high energy  cosmic ray flux.

Due to their low interaction probability, neutrinos need to interact with a large amount of mat
ter in order to be detected. One of the detection techniques is based on the detection of
EAS in the atmosphere by looking for very inclined young showers. The neutrino events would have a significant electromagnetic component leading to a broad time structure of the detected signal, in contrast
to nuclei-induced showers.

\begin{figure}[t]
\centering
\includegraphics[width=7.5 cm,height=6cm]{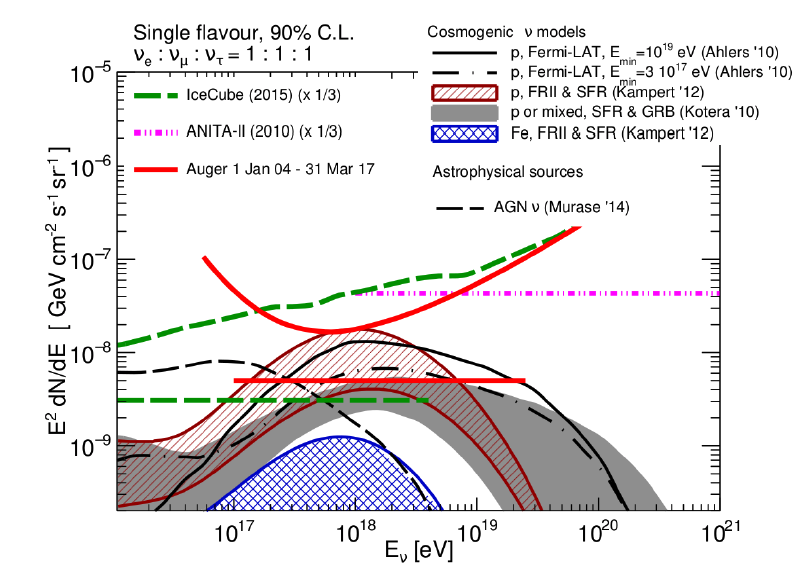}
\includegraphics[width=8 cm,height=5.5cm]{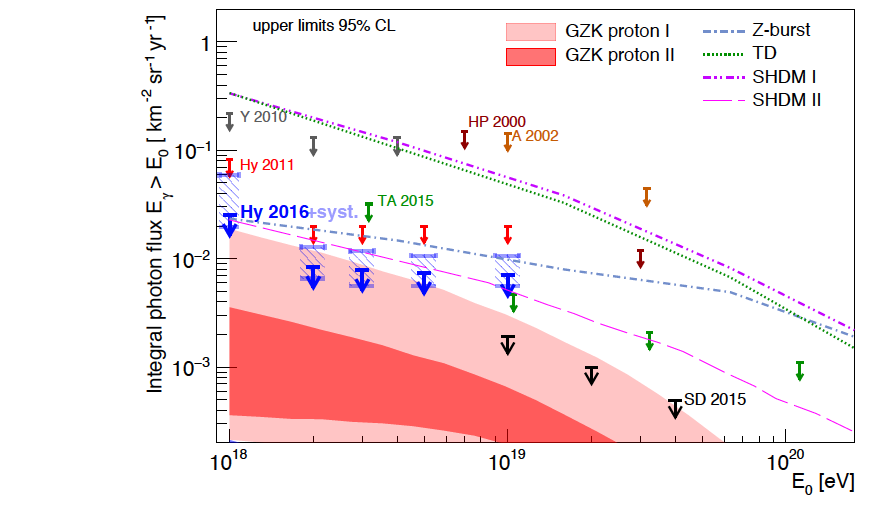}
\caption{Left: Differential (curved red line) and integrated (straight red line) upper limits (90 \% C.L.) from the Pierre Auger Observatory for a diffuse flux of neutrinos in the period 1 Jan 2004 -  31  March 2017. The limits are drawn for a single flavor, assuming equal flavor ratios. Limits from ANITA (magenta dotted dashed line), and IceCube (green dashed line), along with several model predictions are also shown, see~\cite{zas}  for the complete set of references. Right: Upper limits on the integrated photon flux, along with several model predictions. The bold arrows (Hy 2016) correspond to the most recent data analysis in Auger using hybrid events, the blue dashed boxes mark the systematic uncertainties of this study. The limits of previous studies done by Auger,
Hy2011 (red arrows), and SD 2015 (black arrows), the latter one using only SD data, are also shown. The
results from other experiments are also presented.  See~\cite{niechciol,photons}  for the complete set of references. }
\label{fig-neutrino}
\end{figure}

When propagating through the Earth, only the so-called Earth skimming (ES) tau neutrinos may initiate detectable air showers above the ground~\cite{fargion,bertou}.  In this case tau neutrinos may interact within the Earth and produce charged tau leptons which in turn  decay into neutrinos with lower energies. Since the  interaction length for the produced tau lepton is a few tens kilometers at the energy of about 1 EeV, the
leptons produced close to the Earth’s surface may emerge from the ground, decay in air and produce EAS potentially detectable by the surface detector of the Auger Observatory~\cite{zas}. The SD is also sensitive to down-going neutrinos (DG) in the EeV energy range. Down-going neutrinos of any flavour may interact through both charged
(CC) and neutral current (NC) interactions,  producing hadronic and/or electromagnetic showers, see~\cite{zas} for details about the search technique. Up to now, no candidate events have been found by the Auger Observatory to  fulfill  the selection criteria. The absence of neutrino candidate events yields an upper limit on the diffuse flux of neutrinos in the EeV energy range~\cite{zas}, see  Figure~\ref{fig-neutrino} (left).

The Auger Observatory has also set new photon limits with both the hybrid and SD detection methods~\cite{niechciol,photons}. The new limits are compared to previous results and to theoretical predictions in Figure 6 (right). In terms of the photon fraction, the current bound at 10 EeV approaches the percent level while previous bounds were at the 10 percent level. A discovery of a substantial photon flux could have been interpreted as  a signature of top-down (TD) models. In turn, the  experimental limits now put strong constraints on these models.

\subsection{Multi-messenger astronomy }
\begin{figure}[t]
\centering
\includegraphics[width=7.5 cm,height=5.5cm]{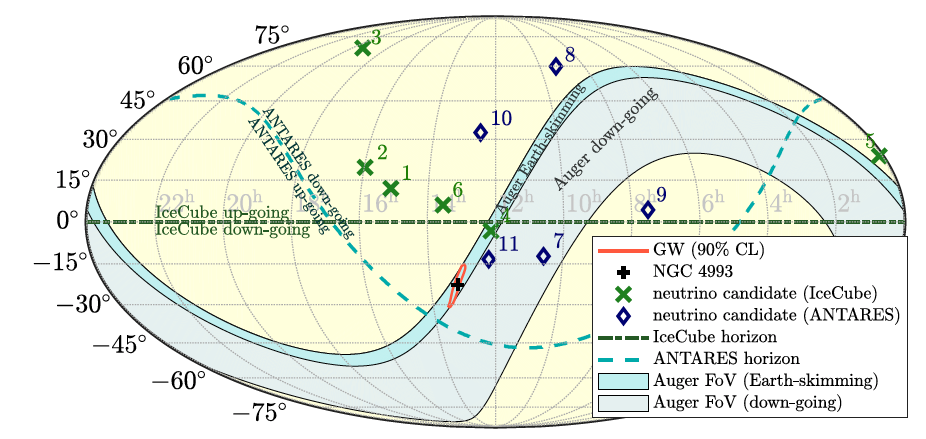}
\includegraphics[width=7.0 cm,height=5.5cm]{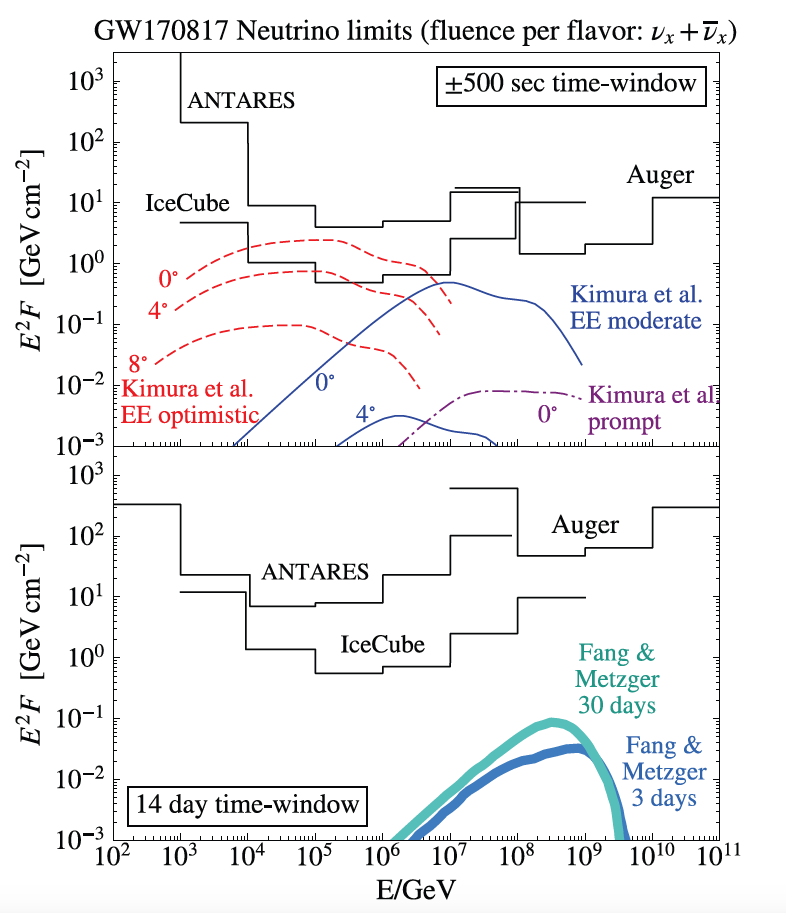}
\caption{ (Left) Sensitive sky areas of ANTARES, IceCube and Auger at the time of the GW170817 event in
Equatorial Coordinates~\cite{augergw1}.  (Right): Upper limits at 90\% C.L. of the neutrino spectral fluence from GW1700817 event for 
a 500s time window (top panel) and in the following 14 days after the trigger (bottom panel)~\cite{augergw1}.
}
\label{fig-gw}
\end{figure}

The Pierre Auger Observatory actively participates in multi-messenger searches in collaboration with other observatories. So far, searches for correlation of  arrival directions of CRs detected by the Telescope Array and Auger collaborations, with  arrival directions of neutrino events detected by IceCube have been performed~\cite{auger-icecube}. No indications of  correlations  at discovery level were found. The detection of the first Gravitational Wave
(GW) transient GW150914 on September 14, 2015, by the Advanced LIGO detectors opened a new
era in multi-messenger astronomy~\cite{ligo}, enhancing the participation of many other experiments, of
which  the Pierre Auger Observatory is also a part. Auger performed neutrino searches in coincidence with the gravitational wave events
 GW150914, GW151226, LVT15012~\cite{augergw1}, and GW170817/GRB 170817A~\cite{augergw2}.
 The GW170817/GRB 170817A~\cite{gw2}, a gravitational wave event detected on August 17, 2017, was later observed as a short gamma-ray burst by the Fermi-GBM and INTEGRAL.  This event was caused by the merging of a binary of neutron stars in the host galaxy NGC4993 at a distance of $40$ Mpc, the closest  gravitational event detected so far. Neutrino searches related 
to this event were carried out by the three most sensitive neutrino observatories IceCube, Antares and Auger. The sky map of these neutrino searches is shown in Figure~\ref{fig-gw}. In Auger, the whole   $\pm 500$ s time window was observed in the ES channel field of view. No inclined showers passing the ES channel selection were detected during this period. Assuming neutrinos are emitted steadily during this period, with an energy spectrum of $E^{-2}$~\cite{augergw1},
 the non-detection of candidates allows us to put limits to neutrino fluence, see Figure~\ref{fig-gw} (right). In the following 14
days, searches were done both in the ES and DG channels.
No neutrino candidates were found in this time window either. No significant counterpart
 was found in any of the searches with any of the observatories, a result which is compatible with
the expectations of a GRB observed off-axis, Figure~\ref{fig-gw} (right).
\subsection{Auger Prime} 
Taking stable data since 2004, the Pierre Auger Collaboration has published many results about 
properties of UHECRs. Among the latest results at the highest energies using SD data, the extragalactic origin of cosmic rays with E > 8 EeV~\cite{science} was
established. However, to identify the
UHECR sources, it is crucial to determine the nuclear mass composition in the flux suppression
energy region. Unfortunately, the low duty cycle of the FD does not allow collecting a significant data sample
for energies above $10^{19.6}$ eV~\cite{augerXmax}. Several other mass composition analyses using the SD were performed 
by the Pierre Auger Collaboration, but these suffer from larger systematic uncertainties due
to the uncertainties in the assessment of the muon content of the shower using the water-Cherenkov
detectors. Thus it is  also   difficult to build a consistent picture of the origin of UHECRs. 

\begin{figure}[th]
\centering
\includegraphics[width=12.0 cm]{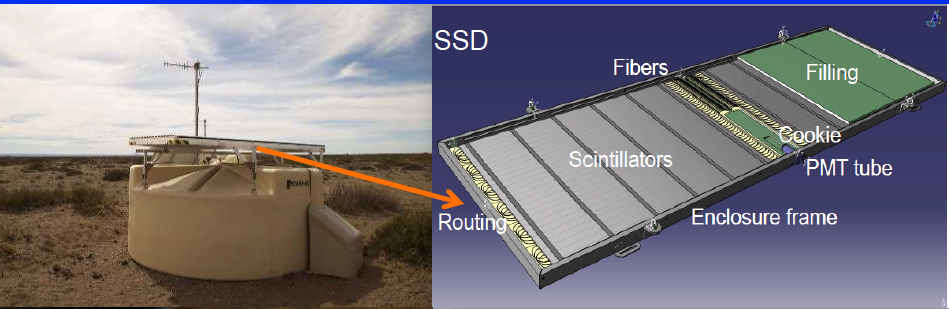}
\caption{The picture of one upgraded station  (left)  and the layout of the Surface Scintillator Detector  (right).
}
\label{fig-prime}
\end{figure}

To address such challenges, the Pierre Auger Observatory is currently undergoing a major
upgrade phase, called AugerPrime~\cite{augerprime}. The key upgrade element is  the installation of a 4 m$^{2}$ plastic scintillator detector on top of each  of the 1660 water-Cherenkov stations (Figure~\ref{fig-prime}), enabling a better discrimination between the electromagnetic and
muonic components of the shower. Additionally, the duty cycle of the fluorescence telescopes will
be extended, allowing a direct determination of the depth of shower maximum with increased
statistics at the highest energies. The electronics of the SD stations is being upgraded
to obtain an increased sampling rate and a better timing accuracy, as well as a higher dynamic range,
 allowing a better reconstruction of the geometry of the showers.
 
The first step for ongoing upgrade is the AugerPrime Engineering Array (EA)~\cite{augerprime2}, which consists  of 12 upgraded detector stations already operational  since 2016.  With this setup we have verified the basic functionality of the detector design, the linearity of the scintillator signal, the calibration procedures and operational stability.  The construction of AugerPrime is expected to be finished in 2019 and it will be followed by data-taking   until 2025.
 \section{Conclusions} The measurements performed at the Pierre Auger Observatory  indicate a change in the nature of cosmic rays at around 3 EeV and show a
change in the shape of the energy spectrum and of the elongation rate. These measurements add support
to the hypothesis that an extragalactic component of mixed composition starts to dominate in this energy range. The
recently observed dipole anisotropy has an orientation which indicates an extragalactic origin of UHECRs.
The near future particle accelerator results and AugerPrime will constrain the hadronic interaction models so that the interpretation of the evolution of the shower maximum with energy will be more conclusive.  The photon limits exclude most of the top-down scenarios  above 2 EeV, and the neutrino limits, if no neutrino is observed, will improve by more than an order of magnitude. These determinations, together with the arrival directions and mass composition analysis will help in solving the puzzle of the origin of the highest energy cosmic rays.
 
%%%%%%%%%%%%%%%%%%%%%%%%%%%%%%%%%%%%%%%%%%

%%%%%%%%%%%%%%%%%%%%%%%%%%%%%%%%%%%%%%%%%%
%\funding{The successful installation, commissioning, and operation of the Pierre Auger Observatory would not have been possible without
% the strong commitment and effort from the technical and administrative staff in Malargüe, 
 %and the financial support from a number of funding agencies in the participating countries, listed at %https://www.auger.org/index.php/about-us/funding-agencies.}

%%%%%%%%%%%%%%%%%%%%%%%%%%%%%%%%%%%%%%%%%%
\acknowledgments{The successful installation, commissioning, and operation of the Pierre Auger Observatory would not have been possible without the strong commitment and effort from the technical and administrative staff in Malargue,and the financial support from a number of funding agencies in the participating countries,listed at  https://www.auger.org/index.php/about-us/funding-agencies. In particular we want to acknowledge support in Poland from National Science Centre grant No. 2016/23/B/ST9/01635.
}

%%%%%%%%%%%%%%%%%%%%%%%%%%%%%%%%%%%%%%%%%%
\conflictsofinterest{The authors declare no conflict of interest} 

%%%%%%%%%%%%%%%%%%%%%%%%%%%%%%%%%%%%%%%%%%
%% optional
\abbreviations{The following abbreviations are used in this manuscript:\\

\noindent 
\begin{tabular}{@{}ll}
UHECRs& ultra high energy cosmic rays \\
CRs & cosmic rays\\
SD& surface detector\\
FD & fluorescence detector\\
ES& earth skimming\\
DG & down-going\\
EAS & extensive air shower
\end{tabular}}

%%%%%%%%%%%%%%%%%%%%%%%%%%%%%%%%%%%%%%%%%%
%% optional

%=====================================
% References, variant A: internal bibliography
%=====================================
\reftitle{References}

% The following MDPI journals use author-date citation: Arts, Econometrics, Economies, Genealogy, Humanities, IJFS, JRFM, Laws, Religions, Risks, Social Sciences. For those journals, please follow the formatting guidelines on http://www.mdpi.com/authors/references
% To cite two works by the same author: \citeauthor{ref-journal-1a} (\citeyear{ref-journal-1a}, \citeyear{ref-journal-1b}). This produces: Whittaker (1967, 1975)
% To cite two works by the same author with specific pages: \citeauthor{ref-journal-3a} (\citeyear{ref-journal-3a}, p. 328; \citeyear{ref-journal-3b}, p.475). This produces: Wong (1999, p. 328; 2000, p. 475)

%=====================================
% References, variant B: external bibliography
%=====================================
%\externalbibliography{yes}
%\bibliography{your_external_BibTeX_file}

%%%%%%%%%%%%%%%%%%%%%%%%%%%%%%%%%%%%%%%%%%
%% optional
%\sampleavailability{Samples of the compounds ...... are available from the authors.}

%% for journal Sci
%\reviewreports{\\
%Reviewer 1 comments and authors’ response\\
%Reviewer 2 comments and authors’ response\\
%Reviewer 3 comments and authors’ response
%}

%%%%%%%%%%%%%%%%%%%%%%%%%%%%%%%%%%%%%%%%%%
\end{document}